\documentclass[12pt]{article}

\usepackage{graphicx}
\usepackage{amssymb,latexsym}

\usepackage{xparse}
\usepackage{xspace}
\usepackage{datetime}
\usepackage{amsmath}
\usepackage[labelsep=colon,labelfont=bf]{caption}
\usepackage{tocloft}
\usepackage{pifont}
\usepackage{cite}
\usepackage{color}
\usepackage{collref}
\usepackage[para]{manyfoot}
\usepackage[bottom]{footmisc}
\usepackage{filecontents}
\usepackage[left=3.0cm,right=2.cm,top=2.5cm,bottom=3cm]{geometry}
\usepackage[linktocpage]{hyperref}

\usepackage[T1]{fontenc}

 \usepackage{mathptmx}

\usepackage{suffix}
\usepackage{mathtools}

\DeclarePairedDelimiterX\MeijerM[3]{\lparen}{\rparen}%
{\begin{matrix}#1 \\ #2\end{matrix}\delimsize\vert\,#3}

\newcommand\MeijerG[8][]{%
  G^{\,#2,#3}_{#4,#5}\MeijerM[#1]{#6}{#7}{#8}}

\WithSuffix\newcommand\MeijerG*[7]{%
  G^{\,#1,#2}_{#3,#4}\MeijerM*{#5}{#6}{#7}}

\numberwithin{equation}{section}

\NewDocumentCommand\eqn{mo}{%
  \IfNoValueTF{#2}
     {\[ #1 \]}
     {\begin{equation}\label{#2} #1 \end{equation} \expandafter\newcommand\csname #2\endcsname{\eqref{#2}\xspace}\ignorespaces}
}
\NewDocumentCommand\eqna{mo}{%
  \IfNoValueTF{#2}
    {\begin{align*} #1 \end{align*}}
    {\begin{equation}\label{#2}\begin{split} #1 \end{split}\end{equation} \expandafter\def\csname #2\endcsname{\eqref{#2}\xspace}\ignorespaces}
}
\NewDocumentCommand\twoseqn{momoo}{%
    \IfNoValueTF{#5}
       {\begin{subequations}\begin{align} #1\label{#2} \\ #3 \label{#4}  \end{align}\end{subequations} \expandafter\def\csname #2\endcsname{\eqref{#2}\xspace}\ignorespaces \expandafter\def\csname #4\endcsname{\eqref{#4}\xspace}\ignorespaces}
       {\begin{subequations}\label{#5}\begin{align} #1\label{#2} \\ #3 \label{#4}  \end{align}\end{subequations} \expandafter\def\csname #5\endcsname{\eqref{#5}\xspace}\ignorespaces \expandafter\def\csname #2\endcsname{\eqref{#2}\xspace}\ignorespaces \expandafter\def\csname #4\endcsname{\eqref{#4}\xspace}\ignorespaces}
}
\NewDocumentCommand\threeseqn{momomoo}{%
   \IfNoValueTF{#7}
     {\begin{subequations}\begin{align} #1\label{#2} \\ #3 \label{#4} \\ #5 \label{#6} \end{align}\end{subequations} \expandafter\def\csname #2\endcsname{\eqref{#2}\xspace}\ignorespaces \expandafter\def\csname #4\endcsname{\eqref{#4}\xspace}\ignorespaces \expandafter\def\csname #6\endcsname{\eqref{#6}\xspace}\ignorespaces}
     {\begin{subequations}\label{#7}\begin{align} #1\label{#2} \\ #3 \label{#4} \\ #5 \label{#6} \end{align}\end{subequations} \expandafter\def\csname #7\endcsname{\eqref{#7}\xspace}\ignorespaces \expandafter\def\csname #2\endcsname{\eqref{#2}\xspace}\ignorespaces \expandafter\def\csname #4\endcsname{\eqref{#4}\xspace}\ignorespaces \expandafter\def\csname #6\endcsname{\eqref{#6}\xspace}\ignorespaces}
}
\NewDocumentCommand\fourseqn{momomomoo}{%
   \IfNoValueTF{#9}
     {\begin{subequations}\begin{align} #1\label{#2} \\ #3 \label{#4} \\ #5 \label{#6} \\ #7\label{#8} \end{align}\end{subequations} \expandafter\def\csname #2\endcsname{\eqref{#2}\xspace}\ignorespaces \expandafter\def\csname #4\endcsname{\eqref{#4}\xspace}\ignorespaces \expandafter\def\csname #6\endcsname{\eqref{#6}\xspace}\ignorespaces \expandafter\def\csname #8\endcsname{\eqref{#8}\xspace}\ignorespaces}
     {\begin{subequations}\label{#9}\begin{align} #1\label{#2} \\ #3 \label{#4} \\ #5 \label{#6} \\ #7\label{#8} \end{align}\end{subequations} \expandafter\def\csname #9\endcsname{\eqref{#9}\xspace}\ignorespaces \expandafter\def\csname #2\endcsname{\eqref{#2}\xspace}\ignorespaces \expandafter\def\csname #4\endcsname{\eqref{#4}\xspace}\ignorespaces \expandafter\def\csname #6\endcsname{\eqref{#6}\xspace}\ignorespaces \expandafter\def\csname #8\endcsname{\eqref{#8}\xspace}\ignorespaces}
}

\NewDocumentCommand\newsec{mo}{%
  \IfNoValueTF{#2}
     {\section{#1}}
     {\section{#1}\label{#2} \expandafter\gdef\csname #2\endcsname{\ref{#2}\xspace}\ignorespaces}
}
\NewDocumentCommand\subsec{mo}{%
  \IfNoValueTF{#2}
     {\subsection{#1}}
     {\subsection{#1}\label{#2}\expandafter\gdef\csname #2\endcsname{\ref{#2}\xspace}\ignorespaces}
}
\NewDocumentCommand\subsubsec{mo}{%
  \IfNoValueTF{#2}
     {\subsubsection{#1}}
     {\subsubsection{#1}\label{#2}\expandafter\gdef\csname #2\endcsname{\ref{#2}\xspace}\ignorespaces}
}

\makeatletter
\renewcommand\section{\@startsection {section}{1}{\z@}%
{-6ex \@plus -1ex \@minus -.2ex}%
{2.3ex \@plus.2ex}%
{\bfseries}}
\makeatother
\makeatletter
\renewcommand\subsection{\@startsection{subsection}{2}{\z@}%
                                     {-3.25ex\@plus -1ex \@minus -.2ex}%
                                     {1.5ex \@plus .2ex}%
                                     {\itshape}}
\makeatother
\makeatletter
\renewcommand\subsubsection{\@startsection{subsubsection}{3}{\z@}%
                                     {-3.25ex\@plus -1ex \@minus -.2ex}%
                                     {1.5ex \@plus .2ex}%
                                     {\itshape}}
\makeatother
\makeatletter 
\def\@seccntformat#1{\csname the#1\endcsname.\hspace{4.6pt}} 
\makeatother 

\renewcommand{\appendix}{\appendices}

\newenvironment{acknowledgments}{\vspace{12pt}\begin{center}\textbf{Acknowledgments}\end{center}\vspace{-12pt}}{}

\collectsep{\ding{70}~}
\newcommand{\foot}{\footnote}
\newdateformat{mydate}{\monthname[\THEMONTH] \THEYEAR} 
\mydate{}

\DeclareNewFootnote[para]{E}[roman]
\newcommand{\email}[1]{\footnoteE{\href{mailto:#1}{\texttt{#1}}}}
\newcommand{\emails}[1]{\let\thefootnote\relax\footnotetext{{\texttt{#1}}}}

\makeatletter
	\renewcommand{\abstract}[1]{\def \@abstract {#1}}
	\newcommand{\affiliation}[1]{\def \@affiliation {#1}}
	\newcommand{\preprint}[1]{\def\@preprint {#1}}
	\abstract{}
	\affiliation{}
	\preprint{}
\makeatother

\makeatletter
\def \maketitle {%
	\begin{titlepage}
	          \begin{flushright}
                           \@preprint
                   \end{flushright}
		\begin{center}
			{\Large\bfseries \@title} 
		
			\bigskip\bigskip\bigskip
		
			\@author 
			
			\bigskip
			
			\emph{\@affiliation}
			
                  	\end{center}
			\bigskip

			\noindent\@abstract
			
			\vfill\vfill\vfill\vfill\vfill\vfill\vfill\vfill\vfill\vfill\vfill
			\vfill\vfill\vfill\vfill\vfill\vfill\vfill\vfill\vfill\vfill\vfill

			\noindent\@date
	\end{titlepage}
}
\makeatother

\usepackage{epsfig}
\usepackage{graphics}
\usepackage{indentfirst}
\usepackage{amsmath, amsthm}

\let\a=\alpha \let\b=\beta \let\g=\gamma \let\d=\delta \let\e=\epsilon
  \let\th=\theta
 \let\k=\kappa
\let\l=\lambda \let\m=\mu \let\n=\nu \let\x=\xi \let\p=\pi 
\let\s=\sigma   \let\f=\phi \let\c=\chi \let\y=\psi
\let\vp=\varphi 
\let\w=\omega      \let\G=\Gamma  \let\Th=\Theta 
\let\X=\Xi  \let\S=\Sigma  \let\Y=\Psi
 \let\W=\Omega
\let\la=\label  
  
 \def\bd{\begin{document}} \def\ed{\end{document}}
\def\ds{\documentstyle} \let\fr=\frac \let\bl=\bigl \let\br=\bigr
\let\Br=\Bigr \let\Bl=\Bigl
\let\bm=\bibitem
\let\na=\nabla
\def\tU{{\widetilde U}}
\let\pa=\partial \let\ov=\overline
\def\ie{{\it i.e.\ }}
\newcommand{\be}{\begin{equation}}
\newcommand{\ee}{\end{equation}}
\def\ba{\begin{array}}
\def\ea{\end{array}}

\def\bei{\begin{itemize}}
\def\eei{\end{itemize}}

\def\ben{\begin{enumerate}}
\def\een{\end{enumerate}}

\def\ft#1#2{{\textstyle{{\scriptstyle #1}\over {\scriptstyle #2}}}}
\def\fft#1#2{{#1 \over #2}}
\def\F#1#2{{ F_{#1}^{(#2)} }}
\def\cF#1#2{{ {\cal F}_{#1}^{(#2)} }}

\def\R{{\bf R}}
\def\sst#1{{\scriptscriptstyle #1}}
\def\oneone{\rlap 1\mkern4mu{\rm l}}
\def\e7{E_{7(+7)}}
\def\td{\tilde}
\def\wtd{\widetilde}
\def\im{{\rm i}}
\def\bog{Bogomol'nyi\ }
\newcommand{\ho}[1]{$\, ^{#1}$}
\newcommand{\hoch}[1]{$\, ^{#1}$}
\newcommand{\bea}{\begin{eqnarray}}
\newcommand{\eea}{\end{eqnarray}}
\newcommand{\ra}{\rightarrow}
\newcommand{\lra}{\longrightarrow}
\newcommand{\Lra}{\Leftrightarrow}
\newcommand{\bp}{\tilde \beta^\prime}
\newcommand{\cB}{{\cal B}}
\newcommand{\cO}{{\cal O}}
\newcommand{\vecx}{\vec{x}}
\newcommand{\vecy}{\vec{y}}
\newcommand{\vecp}{\vec{p}}
\newcommand{\vecq}{\vec{q}}
\newcommand{\tr}{{\rm tr} }
\newcommand{\Tr}{{\rm Tr} }
\newcommand{\NP}{Nucl. Phys. }

\newcommand{\cL}{{\cal L}}
\newcommand{\cA}{{\cal A}}
\newcommand{\cT}{{\cal T}}
\newcommand{\cD}{{\cal D}}
\newcommand{\cH}{{\cal H}}

\def\sst#1{{\scriptscriptstyle #1}}
\def\0{{\sst{(0)}}}
\def\1{{\sst{(1)}}}
\def\2{{\sst{(2)}}}
\def\3{{\sst{(3)}}}
\def\4{{\sst{(4)}}}
\def\5{{\sst{(5)}}}
\def\6{{\sst{(6)}}}
\def\7{{\sst{(7)}}}
\def\8{{\sst{(8)}}}
\def\9{{\sst{(9)}}}
\def\p{{\sst{(p)}}}
\def\q{{\sst{(q)}}}
\def\ve{\varepsilon}
\def\vf{\varphi}
\def\F{\Phi}
\def\wg{\wedge}

\def\thb{\bar{\theta}}
\def\Thb{\bar{\Theta}}
\def\barp{\bar{p}}
\def\barq{\bar{q}}
\def\barc{\bar{c}}
\def\bard{\bar{d}}
\def\e{\epsilon}

\def \bi{\bibitem}
\def \la {\label}

\def \l {\lambda}
\def \tl  {{\tilde \l}}
\def \sql {{\sqrt \l}}
\def \adss {$AdS_5 \times S^5$\ }
\newcommand{\rf}[1]{(\ref{#1})}
\def \ov {\over}

\def\Th{\Theta}
\def\vth{\vartheta}
\def\btheta{{\bar\theta}}
\def\ttheta{{{\tilde\theta}}}
\def\bttheta{{{\bar\ttheta}}}
\def\vth{\vartheta}

\def\ra{\rightarrow}
\def\N{{\cal N}}
\def\F{{\cal F}}
\def\uM{\underline{M}}
\def\uA{\underline{A}}
\def\uN{\underline{N}}
\def\uP{\underline{P}}
\def\ua{\underline{a}}
\def\ub{\underline{b}}
\def\uc{\underline{c}}
\def\ud{\underline{d}}
\def\ue{\underline{e}}
\def\uf{\underline{f}}
\def\ui{\underline{i}}
\def\uj{\underline{j}}
\def\uk{\underline{k}}
\def\ual{\underline{\alpha}}
\def\ube{\underline{\beta}}
\def\um{\underline{m}}
\def\un{\underline{n}}
\def\up{\underline{p}}
\def\uq{\underline{q}}
\def\ur{\underline{r}}
\def\us{\underline{s}}

\def\umu{\underline{\mu}}
\def\unu{\underline{\nu}}
\def\ula{\underline{\l}}
\def\uka{\underline{\k}}
\def\usi{\underline{\s}}
\def\urh{\underline{\r}}
\def\cc{\circ}
\def\eqv{\equiv}

\def\ni{\noindent}

\def\Ep{E^{{}^{(+)}}}
\def\Em{E^{{}^{(-)}}}

\def\Mp{M^{{}^{(+)}}}
\def\Mm{M^{{}^{(-)}}}

\def \ha{{1\ov 2}}

\def\r{\rho}

\def\Y{{\rm Y}}
\def\X{{\rm X}}
\def\tY{\tilde{\rm Y}}
\def\tX{\tilde{\rm X}}
\def\dY{\dot{\rm Y}}
\def\dX{\dot{\rm X}}

\def \J {\mathcal{J}}
\def \del {\partial}

\def\dF{\dot{F}}
\def\dG{\dot{G}}
\def\dx{\dot{x}}
\def\de{\dot{e}}
\def\dr{\dot{r}}
\def\dt{\dot{t}}
\def\dth{\dot{\tt}}
\def\df{\dot{\phi}}

\def\ddx{\ddot{x}}
\def\ddt{\ddot{t}}
\def\ddr{\ddot{r}}
\def\ddth{\ddot{\tt}}
\def\ddf{\ddot{\phi}}

\def \E {{\cal E}}
\def \S {{\cal S}}
\def \J {{\cal J}}

\def\ms{\mathcal{S}}
\def\mj{\mathcal{J}}
\def\soj{\fr{\ms}{\mj}}
\def \R {{\bf R}}
\def \om {\omega}
\def \bE {\bar E}
\def \x {{\cal X}}

\def \bi{\bibitem}
\def \la {\label}

\def \l {\lambda}
\def\foot{\footnote}
\def \tl  {{\tilde \l}}
\def \sql {{\sqrt \l}}
\def \adss {$AdS_5 \times S^5$\ }
\def \ov {\over}

\def \varpi {{\rm w}}

\def\thb{\bar{\theta}}
\def\Thb{\bar{\Theta}}
\def\mb{\bar{\m}}
\def\ab{\bar{\a}}
\def\zb{\bar{z}}
\def\psib{\bar{\psi}}
\def\barp{\bar{p}}
\def\barq{\bar{q}}
\def\barc{\bar{c}}
\def\bard{\bar{d}}
\def\e{\epsilon}
\def\wb{\bar{w}}
\def\lb{\bar{\l}}
\def\Jb{\bar{J}}
\def\Nb{\bar{N}}
\def\Zb{\bar{Z}}
\def\pab{\bar{\pa}}

\def\bg{\bar{g}}

\def\At{\tilde{A}}
\def\Bt{\tilde{B}}
\def\Ct{\tilde{C}}
\def\Dt{\tilde{D}}
\def\Et{\tilde{E}}
\def\Ft{\tilde{F}}
\def\Gt{\tilde{G}}
\def\Ht{\tilde{H}}
\def\It{\tilde{I}}
\def\Mt{\tilde{M}}
\def\Rt{\tilde{R}}
\def\St{\tilde{S}}
\def\at{\tilde{a}}
\def\bt{\tilde{b}}
\def\ct{\tilde{c}}
\def\et{\tilde{e}}
\def\ft{\tilde{f}}
\def\gt{\tilde{g}}
\def\mt{\tilde{\mu}}
\def\nt{\tilde{\nu}}

\def\asth{\hat{*}}
\def\phh{\hat{\phi}}

\def\bA{{\bf A}}

\def\ola{\overleftarrow}
\def\ora{\overrightarrow}
\def\alt{\tilde{\a}}

\def\ra{\rightarrow}
\def\Ra{\Rightarrow}

\def\eh{\hat{e}}
\def\eph{\hat{\e}}
\def\ph{\hat{p}}
\def\alh{\hat{\a}}
\def\beh{\hat{\b}}
\def\gah{\hat{\g}}
\def\Fh{\hat{F}}
\def\muh{\hat{\m}}
\def\nuh{\hat{\n}}
\def\thh{\hat{\th}}
\def\dh{\hat{d}}
\def\ih{\hat{i}}
\def\jh{\hat{j}}
\def\kh{\hat{k}}
\def\deh{\hat{\d}}
\def\wh{\hat{w}}
\def\lah{\hat{\l}}
\def\Ah{\hat{A}}
\def\Ch{\hat{C}}
\def\Omh{\hat{\Omega}}

\def\xh{\hat{x}}

\def\ps{\rlap{\, /}\;\,p }
\def\ks{\rlap{\, /}\;\,k }

\def\gym{g_{YM}}

\def\adot{\dot{a}}
\def\bdot{\dot{b}}
\def\bpa{\bar{\pa}}

\def\pr{\prime}
\def\ssk{\medskip}
\def\bsk{\bigskip}
\def\N{\nabla}
\def\clb{\color{blue}}
\def\clr{\color{red}}
\def\clv{\colo{violet}}

\def\lras{\leftrightarrows}

\def\mx{\mathcal{X}}
\def\my{\mathcal{Y}}

\title{
Absorption cross-sections of small quasi-spherical black holes: the massless scalar case  
}

\author{Tatiana Moskalets$\,^{\spadesuit,}$\email{tatyana.moskalets@gmail.com} and Alexei Nurmagambetov$\,^{\spadesuit,\diamondsuit,\clubsuit
,}$\email{ajn@kipt.kharkov.ua}}

\affiliation{
$^{\spadesuit}$ Department of Physics \& Technology, Karazin Kharkov National University, 4 Svobody Sq., Kharkov, UA 61022
Ukraine\\
$^{\diamondsuit}$
Akhiezer Institute for Theoretical Physics of
NSC KIPT,\\
1 Akademicheskaya St., Kharkov, UA 61108 Ukraine\\
and\\
$^{\clubsuit}$
Usikov Institute for Radiophysics \& Electronics,\\
12 Proskura St., Kharkov, UA 61085 Ukraine
}

\abstract{We consider effects of non-uniformity of quasi-spherical small black hole horizons on scattering massless spineless particles in the long-wave approximation. Focusing on 4D flat and AdS neutral black hole backgrounds with conformally spherical geometry of the horizon, we observe the notable differences in compare to the scattering process on the spherically-symmetric black holes. In particular, the absorption cross-section becomes dependent on both, polar and azimuthal, spherical angles, projections of the angular momentum do not keep anymore and the angular momentum operator by itself, though remains quantised, is not quantised in integers. However, within the long-wave approximation, the main conclusion of previously obtained results on scattering on the spherically-symmetric black holes remains the same: the total absorption cross-section is proportional to the area of the black hole. The proportionality coefficient does not depend on the scalar wave frequency in the flat space black hole background, and is dependent on the admissible from the unitarity requirement frequencies in the background of AdS black hole. As a by-product of our studies we establish a quasi-spherical non-static Vaidya-type black hole solutions and outline the relation between real solutions to the elliptic Liouville equation on 2D plane and on two-dimensional sphere.

\ssk

{
PACS numbers:  04.70.-s, 04.30.Db, 03.65.Nk
}
}

\date{\today}

\begin{document}
\maketitle

\tableofcontents

\section{Introduction}

Physics of black holes attracts a lot of attention since the very beginning. The existing difference (``establishment'' vs ``radicals'') in views on a black hole formation, its evolution and the final state of the process reveals the importance of this object in testing the current paradigm in theoretical physics. We are faced with a controversial situation, when astrophysical data give more and more evidence on the existence of supermassive black holes in galaxies nuclei \cite{Schodel:2002py}, \cite{Schodel:2003gy},\cite{Ghez:2003qj},\cite{Cattaneo:2009ub} (see also \cite{Merloni:2015dda} for a recent review), while recent theoretical studies of the black hole formation give more (to one degree or another) arguments in favour of another mechanism of forming ``black stars'', compatible with the required unitarity of the process, but predicting a finite (even short enough) life-time of the black object during the process of their dissolving \cite{Kawai:2013mda},\cite{Mersini-Houghton:2014zka},\cite{Mersini-Houghton:2014cta},\cite{Ho:2015fja},\cite{Kawai:2015uya},\cite{Ho:2015vga} (see also \cite{Barcelo:2014cla},\cite{Barcelo:2015noa},\cite{Barcelo:2015} in this respect).  This apparent mismatch between theory and experiment motivates further studies of black holes in different viewpoints and directions.

In general, the gravitational collapse is not necessarily restricted to the spherically symmetric consideration. A notable example of a quasi-spherical gravitational collapse was proposed by Szekeres more than 40 years ago \cite{Szekeres:1974ct},\cite{Szekeres:1975dx}. From the point of view of the astrophysical black holes formation this scenario is more realistic in many aspects. Independently on the selected framework, the black hole, once forming, becomes invisible for an external observer. Therefore the main portion of information on the black hole structure can be obtained indirectly, in particular, in studying cross-sections of incident particles, scattering on the black hole. Study of scattering on small black holes with the quasi-spherical geometry of the horizon is the main objective of the paper.

Our consideration is based on the results of our previous paper \cite{Moskalets:2014hoa}, where we 
established new solutions for AdS\(_4\) black holes (BHs) with non-uniform horizons. 
The novelty of the solutions is determined by a special conformal factor obeying the elliptic Liouville equation, hence called the Liouville mode. Our solutions are similar to that of \cite{Szekeres:1974ct},\cite{Szekeres:1975dx} in the main respect, viz., abandoning the strong symmetry group. In fact, one may consider the solutions of \cite{Moskalets:2014hoa} as a partial form of the general Szekeres metric, extended to the black hole solutions to the Einstein equation for AdS space without matter.

It is quite natural to interpret the Liouville mode \cite{Moskalets:2014hoa}
as the distribution function of the spatial inhomogeneity on the horizon virtual surface that may be regarded as a hair of the BH (see Appendix A in the end of the paper and \cite{Moskalets:2015xxa} for more details). Accepting scenarios \cite{Kawai:2013mda},\cite{Mersini-Houghton:2014zka},\cite{Mersini-Houghton:2014cta},\cite{Ho:2015fja},\cite{Kawai:2015uya},\cite{Ho:2015vga} the (near) horizon quasi-spherical surface becomes the real surface of the location of the black hole's forming matter. Then the black hole Liouville hair turns into the real distribution function of the collapsed matter over the angles. To give more evidence in favour of this claim we extend our original black hole solutions \cite{Moskalets:2014hoa} to the non-static Vaidya-type metric \cite{Vaidya:1951zz},\cite{Vaidya:1951zza}, which solves the Einstein equation with the relativistic dust matter source (see Appendix A).

Because any hair contains a portion of information on the BH, one may wonder what the type of information could the Liouville mode contain? And how the non-uniformity, prompted by the spatial inhomogeneity distribution, effects the course of physical processes near the BH horizon?\footnote{In the context of the AdS/CFT correspondence similar questions may be addressed to constructing the non-gravitational side of the duality. For instance, one may wonder how to take into account effects of the space-time inhomogeneity on the CFT side. (A. N. thanks Stam Nicolis for rising this point.)}

If the first of the posed questions admits answering within the BH thermodynamics \cite{Moskalets:2015xxa} (that is, the non-trivial Liouville mode differentiates a stationary black hole state with non-maximal entropy \cite{Moskalets:2014hoa} from the very state of the black hole with the maximum possible amount of ``holographic'' information and with the uniform horizon), the second question is still unanswered. It motivates us to make a step towards filling this gap.

Specifically, we study a feedback of the non-uniformity of the small quasi-spherical BHs horizons on the scattering process of particles in the black hole backgrounds. We consider scattering of massless scalars on two different types of quasi-spherical black holes: the Schwarzschild black holes in four-dimensional Minkowski and AdS space-times. Scattering on the flat space black holes is of potential interest for astrophysics, while solving the same problem in the AdS case is motivated by numerous applications of the AdS/CFT correspondence in condensed matter physics. The results previously obtained on scattering scalars on the spherical BHs (see, e.g., \cite{Unruh:1976fm},\cite{Sanchez:1976},\cite{Sanchez:1977si},\cite{Sanchez:1977vz},\cite{Andersson:1995vi},\cite{Das:1996we},\cite{Harmark:2007jy},\cite{Benone:2014qaa},\cite{Sakalli:2016abx} and monograph \cite{FHM2009}) sharply indicate essential simplification of the task by choosing the appropriate approximation. Here we take the long-wave approximation, specified by two relations, $\w \ll T_H$ and $\w r_+ \ll 1$, between frequencies of the scalar waves and the Hawking temperature/the Schwarzschild radius of the black holes. Also we focus on scattering on small black holes. That is, $r_+ \ll 1$ for Minkowski space and $r_+ \ll 1/\k$ for AdS$_4$ with the characteristic length $1/\k$.

As a result, we observed the following notable differences in the scattering process on the black holes with the quasi-spherical (conformally spherical in precise\footnote{Similarly to the conformally flat metric $ds_{M^2}^2=e^{\Phi(x,y)}(dx^2+dy^2)$ the part $ds^2_{{\cal M}^2}=e^{\chi(\th,\vf)}(d\th^2+\sin^2\th d\vf^2)$ of the whole BH background metric (cf. \rf{BHsol}) is called conformally spherical.}) horizon in compare to the well-known case of the spherically-symmetric black holes \cite{Unruh:1976fm}\;--\cite{FHM2009}:
\begin{itemize}
\item
The presence of the special conformal factor (the Liouville mode) in the BH background metric breaks the spherical symmetry of the Schwarzschild solution. In particular, $k_\vf=\pa/\pa \vf$ is not {a} Killing vector, hence the projection of the total angular momentum on a selected axis does not keep anymore. Moreover, the conformally-spherical structure of the solutions considered here does not preserve the total angular momentum value as well. In effect, we can not separate the contributions of different partial waves to the absorption cross-section and the final expression for the cross-section (differential or total) is dependent on two (polar and azimuthal) spherical angles.
\item
Except of the monopole number, multipole numbers, being integers for the spherically symmetric scattering problem, do not take their values in integers anymore. Their values are determined by the integral average over the angles from the product of two spherical harmonics, weighted by the exponent of the Liouville field. Speaking quantum-mechanically, the total angular momentum spectrum is quantized, but not in integers. This fact prompt us to consider the partial waves with general positive values of the angular momentum azimuthal number. 
\item
{
Frequencies of the scalar perturbations over the AdS black hole background form a discrete spectrum, which is infinity-degenerated for every value of the multipole number. This is the direct consequence of unitarity of the approach. The main contribution to the absorption cross-section comes, within the long-wave and small black holes approximations, from the lowest multipole wave. This $s$-wave contribution to the absorption cross-section consists of the sum over all acceptable $s$-wave frequencies, which retain the long-wave and small black hole approximations.

}

\end{itemize}

However, the main conclusion of \cite{Unruh:1976fm}\;--\cite{FHM2009}, which is quite predictable from  the quantum scattering theory, does not change: the total absorption cross-section is proportional to the small black hole area in the  long-wave approximation. The proportionality factor does not depend on the scalar wave frequency in the flat space neutral BH background, and is the sum over the admissible frequencies of the massless scalar perturbations over the AdS Schwarzschild BH background, leaving intact the approximations we follow here. These results are the content of the main part of the paper, Sections 2 and 3. The last section contains our summary and discussion of open questions. 

For the reader convenience we collect useful information in Appendices. A by-product of our studies is the extension of a static quasi-spherical BH solution in empty space to the Vaidya-type non-static black hole in the presence of the null dust matter. We also outline a way of {solving the} Liouville equation in spherical coordinates. These two issues are addressed in Appendices A and B. In Appendix C we assemble some useful properties of the Gaussian hypergeometric function employed in our computations.

\section{Massless scalar field in the quasi-spherical Schwarzschild background: separation of variables}
 
Let us review scattering of the massless scalar field on a small neutral black hole in the long-wave approximation. This case has been widely considered in the literature (see e.g. \cite{Unruh:1976fm},\cite{Das:1996we},\cite{Harmark:2007jy},\cite{FHM2009}) for black hole backgrounds with the uniform (spherical-type) horizon. Therefore, specifics of scattering on BHs with the non-uniform (spherically-deformed) horizon will be easily determined by the direct comparison of new and old results. Completing this task involves a number of steps to realisation of which we now turn.

The computation of the absorption cross-section is based on {solving the} Klein-Gordon (KG) equation 
\begin{equation}
\fr1{\sqrt{-g^\4}}\,\pa_\m \left(\sqrt{-g^\4}\,g^{\m\n} \pa_\n\Phi\right)=0\,,
\label{KG}
\end{equation}
on a 4D curved background, defined by the metric tensor $g_{\m\n}$. We choose the background metric with the typical structure of a static black hole solution to the flat/curved space Einstein equation:
\be
ds^2=-f(r)dt^2+h(r)dr^2+g(r)\, e^{\chi(\th, \vp)}\left( d\th^2+\sin^2\th\, d\vp^2 \right).
\la{BHsol}
\ee
Zero of $h^{-1}(r)$ corresponds to the horizon location in the radial coordinate direction. Once  $h^{-1}(r)=0$ allows multiple solutions, the true horizon is located at the maximum value $r_+$. Details on the local geometry of the BH virtual horizon are spread over this simulated surface by means of the distribution function $\chi(\th,\vp)$. If we require the correspondence of \rf{BHsol} to the dynamical metric satisfying the Einstein equation the distribution function can not be arbitrary chosen:  
it obeys the Liouville equation \cite{Moskalets:2014hoa} in the considered case. The explicit form of $\chi(\th,\vp)$ is not important in the subsequent discussion.\footnote{More details on solutions to the Liouville equation on the sphere are in Appendix B.}
Clearly, the trivialisation of the distribution function moves back to the old results \cite{Unruh:1976fm},\cite{Das:1996we},\cite{Harmark:2007jy}.

{It is simpler to solve the KG equation} by use of the standard factorisation ansatz
\be
\Phi(t,r,\th,\vf)=e^{-i\w t} \f_\w(r) \, \Th(\th,\vf)\,,
\la{Phifact}
\ee
that results in splitting equation \rf{KG} into the radial and the angular parts. Explicitly,
\be
\fr1{g\sqrt{fh}}\pa_r(g \sqrt{f h^{-1}}\,\pa_r \f_\w)+\fr1f\w^2\f_\w+\fr{1}{g} \f_\w \left(e^{-\chi}\,\fr{\triangle_{\th,\vf} \Th}{\Th}\right)=0\,,
\la{KGsep}
\ee
and the complete separation of variables is achieved by the following requirement for the angular part of \rf{KGsep}:
\be
e^{-\chi}\,\fr{\triangle_{\th,\vf} \Th}{\Th}=C\,.
\la{KGang}
\ee
Here $C$ is the separation constant and $\triangle_{\th,\vf}$ is the angular part of the Laplacian in spherical coordinates:
\be
\triangle_{\th,\vf}=\fr1{\sin\th}\pa_\th\left( \sin\th \, \pa_\th \right) +\fr1{\sin^2 \th}\pa^2_\vp\,.
\la{Lapang}
\ee
Once the distribution function of angle variables $\chi(\th,\vf)$ is chosen to be constant we get the standard solution to \rf{KGang}:
\be
\Th(\th,\vf)=\sum_{lm}\,\a_{lm}\,Y_{lm}(\th,\vf),\qquad C=-e^{\chi}\,l(l+1)=\mathrm{const}\,.
\la{Kgangsolstand}
\ee
This choice brings us to the spherically symmetric background back, when the separation constant solely depends on the spherical harmonic degree $l\ge 0$ 
and does not depend on its order $|m|\le l$. Put it differently, scattering in the spherically symmetric background depends on the polar angle $\th$ and does not depend on the azimuthal angle $\vf$ \cite{Friedrich13}. Consequently, in addition to the conserved angular momentum $\vec{l}$, we have another integral of motion $l_z$ {which} makes possible {further separating} the angle variables to $\Th(\th,\vf)=F(\th)M(\vf)$.

In general situation with a non-trivial Liouville mode $\chi(\th,\vf)\ne {\mathrm{const}}$ we expect the dependence of the separation constant on both $(l,m)$ {that} means the dependence of scattering outcomes on two angles $(\th,\vf)$. Since the spherical harmonics form a complete basis on the sphere  we can expand both $\chi(\th,\vf)$ and $\Th(\th,\vf)$ in $Y_{lm}(\th,\vf)$ and make use of the spherical harmonics orthogonality to get
\be
C_{L,M}=-L(L+1) \left[\int\,d\W_2 \,e^\chi Y_{LM}Y^*_{LM} \right]^{-1}\equiv -\n(\n+1).
\la{Cnun}
\ee
Note that $Y_{LM}(\th,\vf)$  is the spherical harmonic with {\it fixed integers} $L\ge 0$ and $|M|\le L$. But in general $\n$ from \rf{Cnun} is not an integer, except $\n=0$. Now it becomes clear that with the dependence on the integer value of the spherical harmonic degree $L$ the separation constant $C_{L,M}$ also depends on the integer value of the spherical harmonic order $M$. The angular momentum is not conserved on the background  \rf{BHsol} (with the non-trivial Liouville mode $\chi(\th,\vf)$) and its projection on any preassigned axis is no longer an integral of motion. Therefore, we can not further factorize the angle dependence in \rf{Phifact}. Also we have to keep the contribution of all partial waves, now indexed by the azimuthal number $\n$, into the total cross-section, picking out the main contribution in the end.

With the separation constant \rf{Cnun} the radial part of the KG equation \rf{KGsep} becomes a Schr\"odinger-type equation for the stationary state $g^{1/2}(r) \f_\w(r)$ 
\be
\left[\pa^2_x +\w^2-V(x(r))\right](g^{1/2} \f_\w)(r)=0\,.
\la{KGSchro}
\ee
Writing down \rf{KGsep} to the Schr\"odinger-type equation \rf{KGSchro} we have introduced the tortoise coordinate (see, e.g., \cite{Unruh:1976fm}, \cite{Das:1996we}, \cite{Harmark:2007jy} or early papers \cite{Wheeler:1955zz},\cite{Regge:1957td},\cite{Finkelstein:1958zz})
\be
dx=\fr{dr}{ \sqrt{f(r) h^{-1}(r)}}\,.
\la{dtau}
\ee
Then the explicit form of the ``potential'' $V(x)$ turns out to be as follows:
\be
V(x(r))=\n(\n+1) \fr{f}{g}+\fr1{g^{1/2}}\pa^2_x \left(g^{1/2}\right)\,.
\la{V[x]}
\ee

\ssk
\section{Grey-body factors and absorption cross-sections of small quasi-spherical Schwarz\-schild black holes 
}

From now on we identify the background \rf{BHsol} with metric of the Schwarzschild black hole in Minkowski and AdS spaces:
\be
h(r)=f^{-1}(r),\qquad  g(r)=r^2\,,
\la{spechg}
\ee
\be
f(r)=1+\k^2 r^2-\fr{r_+}{r}\left(1+\k^2 r^2_+\right)
\quad  \mathrm{ ( \k =0 \,\,Minkowski; \,\k \ne 0 \,\,AdS)}, 
\la{specAdS}
\ee
where $\k$ is the inverse characteristic length of the AdS space. In both cases eq. \rf{KGSchro} turns into
\be
\left[\pa^2_x+\w^2-V(r) \right]Q_\w(r)=0,\quad Q_\w(r)=\left(\fr{r}{r_+}\right) \f_\w,\quad dx=\fr{dr}{f(r)},
\la{KGx}
\ee
\be
V(r)=\fr{f(r)\,\pa_r f(r)}{r}+\fr{f(r)}{r^2}\,\n(\n+1)\,,
\la{Vdef}
\ee
with the red-shift factors $f(r)$ from \rf{specAdS}. To compute the grey-body factors of the scattering on the black hole waves we have to solve the corresponding equations in three space-time regions, usually referred as the near, the mid and the far zones (see, e.g., \cite{Unruh:1976fm}, \cite{Harmark:2007jy}), patch solutions to each other and compute the transmission and reflections coefficients. This is the standard strategy in solving this task (see, e.g., \cite{Harmark:2007jy} for details).

We start from the vicinity near the black hole horizon. In the near zone, close to the horizon location $r_+$, the effective potential $V(r)$ is negligibly small whatever background we set (Minkowski or AdS BHs). Eq. \rf{KGx} turns into
\be
[\pa_x^2+\w^2]Q_\w(x)=0\,.
\la{KGnz}
\ee
The solution with the standard in-coming wave boundary condition,
\be
Q_\w(x)=A_1 e^{-i\w x}
\la{KGsolnz}
\ee
gives the total flux near the BH horizon with the area $\hat{\W}_2 r_+^2$:
\be
J_{hor}=\hat{\W}_2 r^2_+ \,\fr1{2i}\text{W}\{Q_\w,Q^*_\w\}=|A_1|^2 \w\,\hat{\W}_2 r_+^2\,.
\la{JhorRW}
\ee
Here $W\{\f_1,\f_2\}$ denotes the Wronskian of two linearly independent solutions $\f_{1,2}$ and 
\be
\hat{\W}_2=\int\,d\W_2\,e^{\chi(\th,\vf)}\equiv \int_0^{2\pi}\,d\vf \int_0^\pi \,d\th \,\sin\th\,e^{\chi(\th,\vf)}
\la{hatWdef}
\ee
is the average of exp of the inhomogeneity distribution function over the angles.

The near zone red-shift factor $f(r)$ is well approximated by
\be
f(r)\Big|_{r-r_+ \ll r_+} \approx f'(r)\Big|_{r_+}(r-r_+)
\la{f[r]nz}
\ee
so we can {easily} integrate the tortoise coordinate equation $dx=dr/f(r)$ near the horizon. In the considered cases (Minkowski and AdS) the integration results in
\be
x=r_+ \ln\left(\fr{r}{r_+}-1 \right)+c_1\,.
\la{torcnz}
\ee
Ultimately, within the considered here long-wave approximation $\w r_+ \ll 1$, the near-zone solution to the KG equation \rf{KGsolnz} becomes
\be
Q_\w(r)\approx A_1-A_1 i\w \left(r_+ \ln\left(\fr{r}{r_+}-1 \right)+c_1 \right)\,.
\la{Qnhr}
\ee

Further away from the black hole we arrive at the mid zone, where $V(r) \gg \w^2$. Now the KG equation \rf{KGx} becomes
\be
[\pa^2_x-V(r)]Q_\w=0,
\la{KGmz}
\ee
so the details on the BH background, affecting the shape of $V(r)$, turn out to be important. 
This is the reason to consider different backgrounds (with $\k=0$ and $\k \ne 0$) separately.

\subsection{The grey-body factor of the flat-space Schwarzschild BH}

Computing the effective potential for the flat-space Schwarzschild BH background we arrive at the following
solution to the mid zone equation \rf{KGmz}:
\be
Q_\w(r)=C_1 \,r \,P_\n\left(2\fr{r}{r_+}-1\right)+C_2 \,r \,Q_\n\left(2\fr{r}{r_+}-1\right)\,,
\la{KGmzsol}
\ee
where $P_\n(z)$ and $Q_\n(z)$ are the 1st and the 2nd kind Legendre functions. The near-horizon asymptotics of \rf{KGmzsol} 
\be
Q_\w(r\ra r_+)=C_1\, r_++C_2 \,\fr{r_+}{\G(\n+1)}\left[-\fr12\ln\left(\fr{r}{r_+}-1\right)-\hat{\g}-\hat{\y}(1+\n) \right]
\la{RWsolj0nh}
\ee
is used to fix the undetermined integration constants $C_{1,2}$, linking the solutions in different zones. We choose 
\be
C_1=A_1/r_+,\qquad C_2=2iA_1\w\,\G(1+\n)\,,
\la{C12j0}
\ee
leaving the Euler constant $\hat{\g}$ and the digamma function $\hat{\psi}(z)=\G'(z)/\G(z)$ to fix the undetermined constant $c_1$ in \rf{Qnhr}.

Far away from the BH horizon, at large values of $r$, we come to the far zone. Here the KG equation
\be
\pa^2_r Q_\w(r)+\w^2 Q_\w(r)-\fr{\n(\n+1)}{r^2} \,Q_\w(r)=0
\la{KGfz}
\ee
is solved with
\be
Q_\w (r)=b_1 \sqrt{r} \left(J_{\fr12+\n}(r \w)+i J_{-\fr12-\n}(r \w)\right)+b_2 \sqrt{r}\left(i J_{-\fr12-\n}(r \w)-J_{\fr12+\n}(r \w)\right)\,.
\la{KGfzsol}
\ee
The small $r$ asymptote of \rf{KGfzsol}
\be
Q_\w(r \ra 0)\approx i(b_1+b_2)\, \fr{2^{\n+\fr12} \w^{-\n-\fr12}}{\G(\fr12-\n)}\,r^{-\n}+(b_1-b_2)\,\fr{2^{-\n-\fr12}\w^{\fr12+\n}}{\G(\fr32+\n)}\,r^{\n+1}
\la{KGfzsolr00}
\ee
has to be linked with the large $r$ asymptote of the mid zone solution \rf{KGmzsol} with the integration constants \rf{C12j0}:
\be
Q_\w(r \gg 1)\approx A_1\left[\fr{2^{2\n}}{\sqrt{\pi}}\fr{\G(\n+\fr12)}{\G(\n+1)}\left(\fr{r}{r_+}\right)^{\n+1}+i
\fr{\sqrt{\pi}}{2^{2\n+1}} \fr{\G(1+\n)}{\G(\n+\fr32)} \w r_+ \left(\fr{r}{r_+}\right)^{-\n} \right]\,.
\la{RWsolj0r}
\ee
Comparing \rf{KGfzsolr00} and \rf{RWsolj0r} to each other we get 
\be
(b_1+b_2)=A_1 \fr{\pi^{\fr12}\G(1+\n)\G(\fr12-\n)}{2^{3\n+\fr32}\G(\n+\fr32)\w^{-1/2}}\,(\w r_+)^{\n+1},\,\,\,
(b_1-b_2)=A_1\fr{2^{3\n+\fr12}\G(\n+\fr12)\G(\n+\fr32)}{\pi^{\fr12}\G(\n+1)\w^{-1/2}}\,(\w r_+)^{-\n-1}\,.
\la{b12}
\ee

To check the validity of the constructed solution \rf{KGfzsol} with $b_{1,2}$ from \rf{b12} let us compute the total flux of the ``wave function'' \rf{KGfzsol} over a closed surface of the area $\hat{\W}_2 r^2$:
\be
J_{r \gg 1}=\hat{\W}_2 r^2\fr1{2i}\mathrm{W}\{Q_\w,Q^*_\w\}=\hat{\W}_2 r^2\fr2{\pi} (b_1+b_2)(b_1-b_2)\cos (\pi \n)\,,
\la{Jfzj0}
\ee
which on account of \rf{b12} turns into
\be
J_{r \gg 1}=\hat{\W}_2 r^2 |A_1|^2 \w\,.
\la{Jfzj01}
\ee
In terms of the true wave function $\f_\w=(r_+/r)Q_\w$ (cf. \rf{KGx}) the total flux at large $r$,
\be
J_{r \gg 1}=\hat{\W}_2 r^2 |A_1|^2 \w \left(\fr{r_+}{r}\right)^2=|A_1|^2 \w\, \hat{\W}_2 r^2_+ 
\la{Jfzj01-1}
\ee
coincides with $J_{hor}$ \rf{JhorRW}. Hence, the flux is conserved as it is required by unitarity of a quantum mechanical problem associated with the Schr\"odinger equation (eq. \rf{KGx} in the case).

Following \cite{Harmark:2007jy} we introduce
\be
\g(\w)\simeq 4\,\fr{b_1+b_2}{b_1-b_2}=\fr{4 \pi \,\G(1+\n)^2\, \G(\fr12-\n)}{2^{6\n+2}\,\G(\n+\fr12)\,\G(\n+\fr32)^2}\,(\w r_+)^{2\n+2}\,,
\la{greyj0}
\ee
which is nothing but the grey-body factor for the scalar scattering waves. Clearly, in the long-wave approximation $\w r_+ \ll 1$ the most significant contribution to the grey-body factor comes from the $s$-wave with $\n=0$. For $\n=0$ eq. \rf{greyj0} coincides with the corresponding expression of \cite{Harmark:2007jy} (check eq. (2.40) therein).

\subsection{The absorption cross-section}

The just obtained grey-body factor enters the absorption cross-section $\s=\g(\w) |\cA|^2$, the other factor of which is the square modulus of the scattering amplitude $\cA(\th,\vf)$. Now we turn to the computation of the scattering amplitude.

According to the standard Rayleigh expansion of the plane wave with the wave vector $\vec{k}$ (say, along $z$ direction) 
\be
e^{ikz}=\sum_l\, i^l \left[4\pi (2l+1)\right]^{1/2}\,j_l(kr)\,Y_{l0}(\cos\th)\,.
\la{pwser}
\ee
This result is applicable to the spherically symmetric scattering problem, when the dependence of spherical harmonics $Y_{lm}$ on $m$ is reduced to $m=0$ (i.e., the scattering outcome is independent on the azimuthal angle $\vf$). Here we consider the spherically-deformed case, when the results of scattering depend on both angles, $\th$ and $\vf$. So we can not directly apply eq. \rf{pwser} and have to extend the Rayleigh plane wave expansion to the quasi-spherical case.

Specifically, a monochromatic wave in $z$ direction can be expanded in spherical waves as
\be
e^{ikz}=e^{ikr\cos\th}=\sum_{l,m}\, \cA_{lm}\,j_l(kr)\, Y_{lm}\,.
\la{Rexpgen}
\ee
In \rf{Rexpgen} $\cA_{lm}$ denotes the spherical wave amplitude, $j_l(kr)$ is the spherical Bessel function and $Y_{lm}$ are the spherical harmonics, forming a complete basis in the space of functions on the sphere. Since we are ultimately interested in the computation of the total cross-section, which is measured in the far from the black hole region $kr \ra \infty$, we can use the spherical Bessel function asymptote
\be
j_l(kr) \underset{kr \to \infty}{\longrightarrow} \,\fr1{kr} \sin \left(kr-\fr{l\pi}2 \right)\,.
\la{jasymp}
\ee
Then, picking up the in-coming wave, one gets
\[
e^{ikz}\approx \sum_{l,m}\, i^{l+1}\,\fr{e^{-ikr}}{2kr}\,\cA_{lm}\,Y_{lm}(\th,\vp)
\]
and the orthogonality of the spherical harmonics leads to
\be
\cA_{lm}=2i^{-l-1} kr \int\,d\W_2\,e^{ikr(1+\cos\th)}Y_{lm}\,.
\la{Klm}
\ee
The integrand of \rf{Klm} is a fast-oscillating (for $kr \ra \infty$) function, hence the main contribution to \rf{Klm} comes from $\cos \th \approx -1$, i.e. from $\th=\pi$. Therefore, \rf{Klm} turns into
\be
\cA_{lm}\approx 2i^{-l-1}kr \int\,d\W_2\,Y_{lm}(\pi,\vp)\,e^{ikr(1+\cos\th)}.
\la{Klm1}
\ee
A little inspection of the spherical harmonics table leads to the conclusion on the non-trivial contribution to \rf{Klm1} solely from $Y_{00},Y_{10},Y_{20},Y_{30},\dots$ spherical harmonics. Hence the Rayleigh expansion \rf{pwser} can be applied even in the more general situation of scattering onto the spherically-deformed targets.

Far from the scattering centre the asymptote of the wave function tends to a plane wave, which is a superposition of in-coming and out-coming spherical waves. Selecting the in-coming to the BH horizon wave we get
\[
e^{ikz}\underset{R \to \infty}{=} \fr{e^{-ikR}}{R} \,\cA(\th,\vp)\,.
\]
On the other hand we may use \rf{pwser} with \rf{jasymp}, picking out the in-coming spherical wave again. Thus, far from the scattering core we can equate two different representations of the same plane-wave,
\[
\sum_l \,\fr{i^{2l+1}}{2k}\left[4 \pi (2l+1)\right]^{1/2} \fr{e^{-ikr}}{r} Y_{l0}=\fr{e^{-ikR}}{R} \,\cA(\th,\vp)\,,
\]
to get
\be
\cA(\th,\vp)=\sum_l \,\fr{i^{2l+1}}{2k}\left[4 \pi (2l+1)\right]^{1/2} \fr{R}{r}\, e^{-ikr+ikR}\,Y_{l0}\,.
\la{cA}
\ee
In the result, the total absorption cross-section
\be
\s^{\mathrm{abs}}_T=\int \,d\W_2\, \,\g(\w)|\cA(\th,\vf)|^2
\la{stotdef}
\ee
is equal to
\be
\s^{\mathrm{abs}}_T=\fr{4\pi}{4 k^2}\,\sum_{l,m} \,(2l+1) \,\g(\w,\n(l,m))\int \, d\W_2\, \left(\fr{R}{r}\right)^2 Y_{l0}\,Y^*_{l0}\,.
\la{stotRr0}
\ee
In the background geometry \rf{BHsol} $R^2=e^{\chi} r^2$, so we arrive at the following general expression for the absorption cross-section of the small spherically-deformed Schwarzschild black hole:
\be
\s^{\mathrm{abs}}_T=\fr{4\pi}{4 k^2}\,\sum_{l=0}^{\infty}\sum_{m=-l}^{l} \,(2l+1)\, \g(\w,\n(l,m))\int \, d\W_2\, e^{\chi(\th,\vf)} \,Y_{l0}\,Y^*_{l0}\,.
\la{stotRr}
\ee

The main contribution in the long-wave approximation $\w r_+ \ll 1$
comes from the partial wave with $l=0=\n$. Taking the grey-body factor \rf{greyj0} and the dispersion relation of a massless field $k^2=\w^2$ we finally get
\be
\s^{\mathrm{abs}}_T=r_+^2 \int\,e^{\chi}\, d\W_2=A_{BH}\,,\qquad l=0=\n\,.
\la{stotl0}
\ee
Hence we come to the standard result of the quantum scattering theory \cite{Friedrich13}: the total scattering cross-section  in the low-frequency approximation ($s$-wave scattering) is always proportional to the area of the scattering centre (cf. \cite{Unruh:1976fm}, \cite{Das:1996we} for the absorption cross-section of a spherically symmetric small black hole).

\subsection{The grey-body factor and the cross-section in AdS space}

Let us consider the scattering problem in the curved AdS background defined by \rf{BHsol} with the radial functions from \rf{spechg}, \rf{specAdS}. As we mentioned before, the near zone solution \rf{Qnhr} is universal in both (flat and curved) cases; indeed, the AdS and the flat space red-shift factors are almost the same near the horizon. {As before}, we are interested in the long-wave approximation, $\w r_+ \ll 1$, to the scattering process. However, we have another scale in the case, the AdS inverse length $\k$, relation of which to the BH horizon value $r_+$ will be important in the subsequent calculations.

Recall, the long-wave approximation in thermal field theory is encoded in two relations 
(see \cite{Unruh:1976fm}, \cite{FHM2009}, \cite{Das:1996we}, \cite{Harmark:2007jy}):
\be
\w \ll T_H,\qquad \w r_+ \ll 1\,.
\la{wapprox}
\ee 
For the flat Schwarzschild BH background there is not a difference between them, since the Hawking temperature $T_H$ is measured in $r^{-1}_+$. In the AdS neutral BH background the first inequality of \rf{wapprox} becomes
\be
\w r_+\,\left(1+3\k^2 r_+^2\right)^{-1} \ll 1\,.
\la{wapproxTads}
\ee
Depending on the relation between the two scales -- $\k$ and $r_+$ -- one may determine large, with $\k^2 r^2_+  \gg 1$, black holes, and small black holes with $\k^2 r_+^2 \ll 1$. The latter case is in the focus of the paper. Hence, relations \rf{wapprox} are also equivalent to each other for the small AdS black holes.

In the near zone $r-r_+ \ll r_+$ two red-shift factors \rf{specAdS} are equal to each other in the long-wave approximation \rf{wapprox}. Therefore, eq. \rf{KGnz} and its solution \rf{KGsolnz} (or \rf{Qnhr}) still hold.

Now the mid zone is defined by $V(r) \gg \w^2$ and splits into two regions. The first one is restricted by $r_+ \lesssim r \ll 1/\k$; here the red-shift factor $f(r)$, hence the effective potential $V(r)$, coincides with that in the flat space. The solution to the KG equation \rf{KGmz} in this part of the mid zone is presented by \rf{KGmzsol}. The other region of the mid zone is located at $r_+ < r $, where $\k^2 r^2 \sim 1$ holds. The resulted KG equation (on account of the small AdS BH condition $\k^2 r_+^2 \ll 1$)
\be
\pa_r\left(f(r) \pa_r Q_\w(r)\right)-\fr{1}{r}\,\pa_r f(r) \,Q_\w(r)-\fr{\n(\n+1)}{r^2} Q_\w(r)=0,\quad f(r)=1+\k^2 r^2\,,
\la{KGAdSmz}
\ee
is solved in terms of the hypergeometric functions with
\be
Q_\w (r ) = C_1\, (\k r)^{1+\n}\, {}_2 F_1 \left[\fr{3+\n}2,\fr{\n}2,\fr32+\n;-\k^2 r^2 \right]
\la{QmedzR}
\ee
\[
+C_2 \,
(\k r)^{-\n}\, {}_2 F_1 \left[-\fr{1+\n}2,1-\fr{\n}2,\fr12-\n;-\k^2 r^2 \right]\,.
\]
The undetermined constants of \rf{QmedzR} can be fixed by linking this part of the complete solution with the expression \rf{RWsolj0r}, taking the $r\ra r_+$ limit of \rf{QmedzR} to this end. However, one could skip this step as we will see shortly. 

In the far zone $r \gg r_+$, where the contributions of $\w^2$ and $V(r)$ are of the same order, the KG equation \rf{KGx} becomes
\be
\pa_r\left(f(r) \pa_r Q_\w(r)\right)+\fr{\w^2}{f(r)}\,Q_\w(r)-\fr{1}{r}\,\pa_r f(r) \,Q_\w(r)-\fr{\n(\n+1)}{r^2} Q_\w(r)=0\,.
\la{KGAdSfz}
\ee
The solution to \rf{KGAdSfz} with the red-shift factor of \rf{KGAdSmz}
\[
Q_\w(r) =C_1 \left(1+\k^2 r^2\right)^{\fr{\hat{\w}}2}\,(\k r)^{1+\n} \, {}_2F_1 \left[\fr{\n+\hat{\w}}{2},\fr{3+\n+\hat{\w}}2,\fr32+\n;-\k^2 r^2\right] 
\]
\be
+C_2 \left(1+\k^2 r^2\right)^{\fr{\hat{\w}}2}\,(\k r)^{-\n}\, {}_2F_1 \left[-\fr{1+\n-\hat{\w}}{2},\fr{2-\n+\hat{\w}}2,\fr12-\n;-\k^2 r^2\right]
\la{Qsolfar}
\ee
coincides with \rf{QmedzR} in the $\hat{\w}\equiv \w/\k \ra 0$ limit. Therefore, we can fix coefficients $C_{1,2}$ by linking the small $r$ asymptotics of \rf{Qsolfar} with \rf{RWsolj0r} directly. In the $\k^2 r^2 \ra 0$ limit ${}_2 F_1[a,b,c;0]=1$ for any non-trivial $c$, so \rf{Qsolfar} is
\be
Q_\w =C_1 \,(\k r)^{1+\n}+C_2 \,(\k r)^{-\n}\,.
\la{Qf2n}
\ee
To join \rf{Qf2n} to \rf{RWsolj0r} we set
\be
C_1=A_1\, \fr{2^{2\n}}{\pi^{1/2}}\,\fr{\G(\n+\fr12)}{\G(\n+1)} \,(\k r_+)^{-\n-1},\qquad
C_2= iA_1\,\fr{\pi^{1/2}}{2^{2\n+1}}\,\fr{\G(\n+1)}{\G(\n+\fr32)}\, \w r_+\,(\k r_+)^{\n}\,.
\la{C1C2}
\ee

Though we have gathered the solutions to the AdS KG equation in different zones altogether, we did not succeed yet. The point is to verify the ``wave function'' flux conservation, which guarantees unitarity of the approach. Naively, one could say that at the radial infinity the effective potential \rf{Vdef} tends to zero; the KG equation at infinity coincides with that of the near horizon; hence for the total fluxes we have $J_{hor}=J_\infty$ and unitarity is preserved. However, this line of reasoning does not take into account effects of the boundary of AdS space. A more delicate analysis is required to this end.

Thus, we are interested in solutions to the KG equation at $\k^2 r^2 \ra \infty$ (this limit defines a ``deep far zone'',  located near the boundary of AdS space) and in the near the boundary asymptotics of \rf{Qsolfar}. To find this asymptote we introduce $z=-k^2 r^2$. Then the combination $z/(z-1)$ tends to one in the $\k^2 r^2 \ra \infty$ limit, while $1/z \ra 0$. Using results of Appendix C (eqs. \rf{Fz/z-11}, \rf{Fz/z-12}) with integration constants \rf{C1C2} we get 
\[
Q_\w(r)=A_1 \left(\fr{r}{r_+}\right)\left[\fr{2^{2\n-1}(\n+\fr12)\G(\n+\fr12)^2}{\G(\n+1)\G(\fr{3+\n-\hat{\w}}2)\G(\fr{3+\n+\hat{\w}}2)}\,(\k r_+)^{-\n} \right.
\]
\[
\left.
+\fr{i\pi^2}{2^{2\n+2}\cos (\pi \n)\, (\n+\fr12)}\,\fr{\G(\n+1)}{\G(\n+\fr12)^2 \G(\fr{2-\n+\hat{\w}}2)\G(\fr{2-\n-\hat{\w}}2)} \,\hat{\w}(\k r_+)^{\n+2}\right.
\]
\be
\left.
+i\,\fr{2\cos(\fr{\pi}2(\n- \hat{\w}))\cos(\fr{\pi}2(\n+\hat{\w}))}{2^{2\n} \,3\,\cos(\pi \n) (\n+\fr12)}\,\fr{\G(\n+1)\,\hat{\w}(\k r_+)^{\n+2}}{\G(\n+\fr12)^2}\,\G\left(\fr{3+\n-\hat{\w}}2\right)\,\G\left(\fr{3+\n+\hat{\w}}2\right)\,(\k r)^{-3}\right]\,.
\la{Qdzone}
\ee

Now we have to link the asymptotics \rf{Qdzone} with solution to \rf{KGAdSfz} near the boundary $\k^2 r^2 \ra \infty$. Following \cite{Harmark:2007jy} we introduce more convenient for the far zone $\k r \gg 1$ variable $u=\hat{\w}/(\k r) \ll 1$ in terms of which eq. \rf{KGAdSfz} in the ``deep far zone'' (where $f(r)\approx \k^2 r^2$) looks like
\be
\left[ \pa^2_u+\left(1-\fr{2}{u^2}-\fr{\n(\n+1)}{\hat{\w}^2} \right) \right]\left(\fr{\f_\w}{u}\right)=0\,,
\la{KGudfar}
\ee
and solves with
\be
\f_\w(u)=\hat{C}_1\,u^{3/2} H^{\1}_{3/2}\left(-iu\sqrt{\fr{\n(\n+1)}{\hat{\w}^2}-1}\,\right)+\hat{C}_2\,u^{3/2} H^{\2}_{3/2}\left(-iu\sqrt{\fr{\n(\n+1)}{\hat{\w}^2}-1}\,\right),
\la{KGudfarsol}
\ee
where $H^{(1,2)}_n$ are the Hankel functions of the 1st and of the 2nd kind. Introducing $\tilde{\w}=(1-\n(\n+1)/\hat{\w}^2)$ we get the following expansion at $u \ra 0$:
\be
\f_\w=\sqrt{\fr2{\pi}}\left[i\left(\hat{C}_2-\hat{C}_1\right)\fr1{\tilde{\w}^{3/4}}+\fr12\,i \left(\hat{C}_2-\hat{C}_1\right)\tilde{\w}^{1/4}\,u^2+\fr13 \left(\hat{C}_1+\hat{C}_2\right)\tilde{\w}^{3/4}\,u^3+\dots \right]\,,
\la{phiuexp01}
\ee
which has to be compared with \rf{Qdzone} to fix the undetermined $\hat{C}_{1,2}$. Eqs.  \rf{Qdzone} and \rf{phiuexp01} can be slightly simplified before comparing to each other\footnote{The second term of \rf{Qdzone} can be skipped within the small black hole approximation $(\k r_+)^2 \ll 1$ (the smallest
constant term of \rf{Qdzone} in $\f_\w(r)$ variables). In \rf{phiuexp01} one may omit the term $\sim u^2$ (cf.  eq. (4.7) in \cite{Harmark:2007jy}) due to its subleading ($\sim u^4$) contribution to the asymptotic flux at the boundary \cite{Harmark:2007jy}.}, so in the outcome we have
\be
\hat{C}_1-\hat{C}_2=iA_1\,\tilde{\w}^{3/4}\, \fr{2^{2\n-3/2}\pi^{1/2}(\n+\fr12)\G(\n+\fr12)^2}{\G(\n+1)\G(\fr{3+\n-\hat{\w}}2)\G(\fr{3+\n+\hat{\w}}2)}\, (\k r_+)^{-\n}\,,
\la{C1n}
\ee
\be
\hat{C}_1+\hat{C}_2=iA_1\,\fr{\pi^{1/2} \cos(\fr{\pi}2(\n- \hat{\w}))\cos(\fr{\pi}2(\n+\hat{\w}))}{2^{2\n-1/2} \cos(\pi \n) (\n+\fr12)\,\tilde{\w}^{3/4}}\,\fr{\G(\n+1)\,\G\left(\fr{3+\n-\hat{\w}}2\right)\,\G\left(\fr{3+\n+\hat{\w}}2\right)
}{\G(\n+\fr12)^2}\,\hat{\w}^{-2}(\k r_+)^{\n+2}\,.
\la{C2n}
\ee

Now let's compute the total flux on the radial infinity. In the ``deep far zone'' one can employ $\pa_x=-\w \pa_u$ ($\pa_x$ is the derivative over the tortoise coordinate $x$, see \rf{dtau}), hence
\be
J_\infty=-\hat{\W}_2\, r^2\,\fr{\w}{2i}\,\text{W}\{\f_\w(u),\f^*_\w(u)\} \,=\hat{\W}_2\, r^2 \,\fr{\w}{\pi}\,\left((\hat{C}_1-\hat{C}_2)(\hat{C}_1^*+\hat{C}_2^*)+\text{h.c.}\right) u^2\,,
\la{Jdfz}
\ee
where we have used the solution \rf{phiuexp01} and have neglected the subleading contribution to $J_\infty$. On account of \rf{C1n}, \rf{C2n} and $u=\hat{\w}/(\k r)$ 
\be
J_\infty=|A_1|^2\hat{\W}_2\,\w r_+^2 \,\fr{\cos(\fr{\pi}2(\n+\hat{\w}))\cos(\fr{\pi}2(\n-\hat{\w}))}{\cos(\pi \n) }=\fr{\cos(\fr{\pi}2(\n+\hat{\w}))\cos(\fr{\pi}2(\n-\hat{\w}))}{\cos(\pi \n) }\, J_{hor}\,.
\la{Jinfn}
\ee
As a result, the flux conservation, which is equivalent to keeping unitarity within the approach, requires the following restriction on the frequencies spectrum:
\be
\fr{\cos(\fr{\pi}2(\n+\hat{\w}))\cos(\fr{\pi}2(\n-\hat{\w}))}{\cos(\pi \n) }=1\quad \leadsto \quad \fr{\cos(\pi\hat{\w})}{\cos(\pi \n)}=1\,.
\la{AdSunit}
\ee
Therefore, we arrive at the following admissible spectrum of frequencies for the massless scalar modes in the Schwarzschild AdS BH background:
\be
\hat{\w}=\n+2 n,\,\,\, n\in \mathbb{Z}\,.
\la{AdSwspec}
\ee
We specially note that the frequencies $\hat{\w}$ are restricted to $\hat{\w}>0$ and $\hat{\w} \in \mathbb{R}$.

To compute the grey-body factor and the total absorption cross-section we introduce (see \cite{Harmark:2007jy} for details)
\be
z(\hat{\w})=\fr{\hat{C}_1-\hat{C}_2}{\hat{C}_1+\hat{C}_2}\,,
\la{zdef}
\ee
in terms of which the grey-body factor $\g(\hat{\w})$ looks like\footnote{The flat space-time grey-body factor \rf{greyj0} was computed exactly in the same way: eq. \rf{greyAdSn} results in \rf{greyj0} with $z=(b_1-b_2)/(b_1+b_2)$ and $b_{1,2}$ from \rf{b12}.}
\be
\g(\hat{\w})=1-\left|\fr{1-z}{1+z}\right|^2=\fr{2(z+\bar{z})}{1+z+\bar{z}+|z|^2}\,.
\la{greyAdSn}
\ee
With $\hat{C}_{1,2}$ of \rf{C1n}, \rf{C2n}
\be
z(\hat{\w},\n)=2^{4\n-2}\,\fr{(\n+\fr12)^2\,\G(\n+\fr12)^4}{\G(\n+1)^2\G(\fr{3+\n-\hat{\w}}2)^2\G(\fr{3+\n+\hat{\w}}2)^2}\,\left(1-\fr{\n(\n+1)}{\hat{\w}^2}\right)^{3/2}
\,\hat{\w}^{2} \,(\k r_+)^{-2\n-2}\,.
\la{znj0}
\ee

For $\n \ne 0$ the grey-body factor \rf{greyAdSn} is equal to zero for the pure image values of $z(\hat{\w},\n)$ as well as for the trivial $z(\hat{\w},\n)$. It happens for $\hat{\w}^2\le \n(\n+1)$, $\n \ne 0$, or taking \rf{AdSwspec}, for
\be
\n \le \fr{4n^2}{1-4n},\quad n \in \mathbb{Z}, \,\, n\ne 0\,.
\la{nNo}
\ee
With $n>0$ this inequality is not realised (the positivity of $\n$ follows from its definition \rf{Cnun}), hence the grey-body factor is non-trivial for the positive values of $n$. However, possible values of $n$ are bounded from above by the maximal value $n_{max}$ for which $(\n+2n_{max}) \k r_+ \ll 1$ still holds.

For a negative $n$ possible values of $\n$ and $n$ are restricted within the long-wave and the small black hole approximations:
\be
\w r_+=\left(\n-2|n|\right) (\k r_+) \ll 1,\quad n<0,\,\,\,n\in \mathbb{Z}\,.
\la{approxAdS}
\ee
At a first glance this restriction still leaves a wide range of the admissible values of $\n$ and $n$ 
\be
\n<\fr{4n^2}{1+4|n|},\qquad n<0,\,\,\, n\in \mathbb{Z}
\la{nYes}
\ee
which leads to the non-trivial grey-body factor. However, the positivity of $\hat{\w}$ also requires $\n-2|n|>0$, hence
\be
2|n|<\n<\fr{4 n^2}{1+4|n|}\,.
\la{nurestr}
\ee
But in its turn the inequality \rf{nurestr} results in
\be
2|n|+8n^2 < 4n^2\,,
\la{nrestr}
\ee
which can not be satisfied for arbitrary negative integer $n$.

The grey-body factor \rf{greyAdSn} of the permitted higher multipole waves is easy to compute by use of \rf{znj0} and the spectrum of the absorbed waves $\hat{\w}=\n+2n$, $n \in \mathbb{Z}^+$, $n<n_{max}$. It leads to
\be
\g(\n,n)\simeq \fr{(\n+2n)}{2^{4\n-4}(4n^2+\n(4n-1))^{3/2}}\,\fr{\G(1+\n)^2\G(\fr32-n)^2\G(\fr32 +\n+n)^2}{(\n+\fr12)^2\G(\n+\fr12)^4}\,(\k r_+)^{2\n+2}\,.
\la{gdAdSmult}
\ee
Clearly, the small black hole approximation results in the subleading contributions of the higher multipole waves (with $\n\ne 0$) to the absorption cross-section.

For the $s$-wave scattering the coming from \rf{znj0} expression for $z(\hat{\w},0)$  
\be
z(\hat{\w},0)=\fr{\pi}{2^2}\,\fr{\G(\fr32)^2}{\G(\fr{3+\hat{\w}}2)^2 \G(\fr{3-\hat{\w}}2)^2}\,\fr{\hat{\w}^2}{(\k r_+)^2}
\la{z0j0}
\ee
coincides with the corresponding expression in \cite{Harmark:2007jy} (cf. eq. (4.38) therein). Small enough values of $\hat{\w}$ {keep} unitarity with a continuous spectrum of values (see eq. \rf{AdSunit}). The critical frequencies of the spectrum \cite{Harmark:2007jy} correspond to $\g(\hat{\w}_c)=0$ and $\g(\hat{\w}_c)=1$.\footnote{It follows from \rf{greyAdSn} for \rf{z0j0} that the grey-body factor
\[
\g(\w)\approx 4\fr{(\w r_+)^2}{(\hat{\w}^2+(\k r_+)^2)^2}
\]
takes just two values within the considered approximations: zero (for all non-critical $\hat{\w}$) and one.
}

The complete $s$-waves reflection ($\g(\hat{\w}_c)=0$) would be occurred at $\hat{\w}_c=2m+3$, $m \in\mathbb{Z}$, $m \ge -1$.\footnote{In the absence of any restriction on $\hat{\w}$ these critical frequencies correspond to the AdS$_4$ normal modes (see, e.g.,\cite{Burgess:1984ti},\cite{Konoplya:2002zu},\cite{Cardoso:2003cj},\cite{Natario:2004jd},\cite{Konoplya:2011qq}).} But these values of the critical frequencies violate the unitarity requirement $\hat{\w} \ll 1$. Other critical frequencies correspond to the $s$-wave transparency of the BH potential (complete transmission); it occurs for $z(\hat{\w}_c,0)=1$, hence \cite{Harmark:2007jy} for 
\be
\hat{\w}_c=\k r_+\,.
\la{hwcrit}
\ee
Note that the critical frequency value \rf{hwcrit} satisfies both the long-wave and the small black hole approximations.

If we relax the requirement of small frequencies $\hat{\w}$ (indeed, within the taken here approximations large enough values of $\hat{\w}$ are still possible), the unitarity requirement \rf{AdSunit} picks out
(cf. \rf{AdSwspec})
\be
\hat{\w}=2n \quad \leadsto \quad \w=2n\k,\quad n\in\mathbb{Z},\quad n>0\,.
\la{AdSwspec0}
\ee
We have
\be
z(\hat{\w})=\fr{\hat{\w}^2}{(\k r_+)^2}\,\fr{1}{(4n^2-1)^2}\equiv \fr{4n^2 }{(\k r_+)^2(4n^2-1)^2} \,,\qquad n=1,2,\dots
\la{zunit}
\ee
and
\be
\g(\hat{\w})=4\,\fr{4n^2}{(4n^2-1)^2}\,\fr{1/(\k r_+)^2}{\left(1+\fr{4 n^2}{( \k r_+)^2 (4n^2-1)^2}\right)^2}\,.
\la{greyun}
\ee
Equations \rf{zunit}, \rf{greyun} present the exact results of substitution of \rf{C1n}, \rf{C2n} with $\n=0$ in \rf{zdef}, then in \rf{greyAdSn}. In the small black hole approximation $(\k r_+)^2 \ll 1$ the expression for the grey-body factor \rf{greyun} simplifies  to
\be
\g(\hat{\w}=2n) \approx 4 \,(\k r_+)^2 \left(2n-\fr1{2n}\right)^2,\qquad n=1,2,3,\dots
\la{greyAdS0s1}
\ee

The absorption cross-section for the massless scalar $s$-waves (see \rf{stotRr}) becomes 
\be
\s^{(n)}_s=4\left(2n-\fr1{2n}\right)^2
\,(\k r_+)^2\,\fr{1}{4 \w^2}\Big|_{\w=2n\k} \int\,e^\chi \,d\W_2=\a \,A_{BH}\,.
\la{tcsAdS0s1}
\ee
\[
\a=\left(1-\fr{1}{(2n)^2}\right)^2,\quad n=1,2,\,\dots
\]
Every admissible frequency of the spectrum makes the independent contribution into the total absorption cross-section, hence
\be
\s^{\mathrm{abs}}_T=\sum_{n=1}^{n_{max}}\,\s^{(n)}_s=A_{BH}\, \sum_{n=1}^{n_{max}}\,\left(1-\fr{1}{(2n)^2}\right)^2\,,
\la{sAdStotj0}
\ee
where we have bounded the sum from above by some maximal value $\w_{max}=2\k n_{max}$ of the $s$-wave frequency for which $\w_{max} \,r_+ \ll 1$ still holds. Then, the total absorption cross-section becomes
\be
\s^{\mathrm{abs}}_T =(n_{max}- \fr12 H_{n_{max},2})A_{BH}\approx n_{max} \,A_{BH} \,.
\la{sAdStotj0f}
\ee
Here $H_{n_{max},2}$ is the second order harmonic number of $n_{max}$ (which is small enough for a large enough value of $n_{max}$).

\bsk

Therefore, we come to the following conclusions:
\begin{itemize}
\item
The higher multipole waves of the massless scalar perturbations over the small AdS$_4$ neutral black hole backrgound give the subleading contribution to the absorption cross-section, hence we can consider them as completely reflected by the BH. The same effect occurs for the scattering waves of non-critical frequencies $\hat{\w} \ll 1$.
\item
The massless scalar $s$-wave with the critical small frequency $\hat{\w}_c=\k r_+$ is completely absorbed by the small AdS$_4$ neutral black hole. Other $s$-wave critical frequencies corresponding to the AdS normal modes are rejected by the unitarity requirement.
\item
For non-small values of $\hat{\w}$ the requirement of unitarity in the scattering process singles out the admissible massless scalar $s$-wave frequencies, correlated with the AdS inverse length as $\w=2n\k$, $1 \le n \le n_{max}$. This spectrum of frequencies is bounded from above with $\w_{max} \,r_+ \ll 1$. Then the total absorption cross-section is equal to the BH area up to some factor, counting the total number of different allowable frequencies of the spectrum which contribute to the cross-section.
\end{itemize}

\section{Summary and open questions}

To summarise, we have studied the scattering problem of the massless scalar perturbation over the neutral static black hole backgrounds with a non-uniformity on the horizons. 
These backgrounds correspond either to an intermediate state of black holes with a non-maximal amount of entropy/information (see \cite{Moskalets:2014hoa},\cite{Moskalets:2015xxa}) before the black hole reduction to the maximum entropy state with the uniform horizon, or to the BH formation in the gravitational collapse with a non-spherically distributed matter (see \cite{Szekeres:1974ct},\cite{Szekeres:1975dx} and Appendix A below). The non-uniformity measure, differentiating the quasi-spherical (conformally spherical in the case) geometry of the horizon from the pure spherical one, is the conformal factor depending on two (polar and azimuthal) angles on the sphere. In the presence of such a deformation we loose the total angular momentum projection conservation law, and the conformally spherical structure of the considered solutions does not support the total angular momentum conservation as well. It in particular means that the partial waves with different angular momentum azimuthal numbers can not be considered independently of each other. As we have established, 
their multipole numbers are not positive integers in the conformally-spherical case and turn into the familiar $l=1,2,3,\dots$ upon trivialisation of the inhomogeneity distribution function. Put it differently, the quantum-mechanical angular momentum operator is still quantized, but generally not in integers. To find the explicit value of the multipole numbers one has to know the real solutions to the elliptic Liouville equation for the distribution function over the spherical angles. We present the solution to this problem in Appendix B.

The working approximations used in the paper are the long-wave approximation and the small black hole approximation. Applying these approximations essentially simplifies the task, and we found the standard and quite predictable result of the quantum scattering theory: the main contribution into the total absorption cross-section comes from the monopole (multipole with zero number) or $s$-wave, and the total cross-section is proportional to the area of the scattering centre. The proportionality coefficient does not depend on the $s$-wave frequency in the Schwarzschild black hole background, but it depends on the frequencies of $s$-waves for the AdS$_4$ Schwarzschild black hole. This difference in the scattering process in flat and curved backgrounds is directly related to the spectrum of the scalar waves frequencies, which is continuous and non-degenerate in flat space, but discrete and infinitely-degenerate in AdS space. As a result, we have to sum over the acceptable in the long-wave approximation $s$-wave frequencies in AdS space instead of taking into account a single frequency $s$-wave contribution to the flat space total absorption cross-section. Clearly, turning back to the trivial Liouville mode, we recover the results of scattering spinless $s$-waves on the spherically symmetric black hole. And vice versa, setting the red-shift factors of the BH solution to one will {result} in scattering spinless $s$-waves on the spherically-deformed scattering centre in the ``soft-wall'' approximation, when a part of the incident radiation is absorbed by the target.\footnote{ The Liouville equation for $\chi(\th,\vf)$ is also required to recover the geodesically complete flat space solution to the Einstein equation, so in the absence of black holes the problem is reformulated as scattering on the flat space background in the conformally-spherical foliation:
\[
ds^2=-dt^2+dr^2+r^2 e^{\chi(\th,\vf)}(d\th^2+\sin^2 \th d\vf^2)\,.
\] 
Now $r_+$ becomes the ``mean radius'' of the scattering center, and the virtual surface of the BH horizon transforms into the target shape. Since our results do not depend on the explicit expression for $\chi(\th,\vf)$, they are directly applicable to such a case.} 
From this point view our computational scheme can be potentially applied to different areas of modern physics, from nuclear physics to biophysics (cf., for instance,\cite{Ziegler:1989},\cite{Palmer:1995},\cite{Mishchenko:2000},\cite{Rother:2014},\cite{Hobolth:2003}). 

Though we have established peculiarities in the scattering process of the massless scalar modes on  small black holes with the non-uniform (quasi-spherical) horizons, the difference in scattering on small black holes with the pure spherical geometry of the horizon and that on the conformally spherical horizon has to be more pronounced with considering the spin waves. Another interesting problem for further studies is related to the non-conservation of the total angular momentum in the conformally-spherical background that may be a source of chaotisation of the scattering particles trajectories. We postpone these and other related topics to future publications.

\bsk
\centerline{\bf Acknowledgements }

\ssk
T.~M. thanks Cosmology Group for kind hospitality during her internship at LPT, Orsay. A.~N. is grateful to S.~Nicolis for rising interesting points on the AdS/CFT and to V.~P.~Berezovoj, Yu.~L.~Bolotin and K.~A.~Lukin for fruitful discussions.

\bsk\bsk\bsk

\section*{Appendix A. Quasi-spherical non-static black hole solutions and the Liouville hair}

\addcontentsline{toc}{section}{Appendix A. Quasi-spherical non-static black hole solutions and the Liouville hair}

\def\theequation{A.\arabic{equation}}
\setcounter{equation}0

In this Appendix we extend the original solutions of \cite{Moskalets:2014hoa} to the non-stationary Vaidya-type metric and give more evidence for handling the Liouville mode as a black hole hair. 

Let's start with the following generalisation of the Schwarzschild metric to a quasi-spherical black hole:
\be
d\tilde{s}^2=-f(r)dt^2+\fr{dr^2}{f(r)}+g(r) e^{\chi(\th,\vf)}(d\th^2+\sin^2 \th\,d\vf^2),
\la{MinksolL}
\ee
\be
f(r)=1-\fr{r_+}{r},\qquad g(r)=r^2,\qquad \triangle_{\th,\vf}\, \chi(\th,\vf)+2e^{\chi(\th,\vf)}-2=0\,.
\la{MinksolL1}
\ee
One may verify that eqs. \rf{MinksolL}, \rf{MinksolL1} solve the Einstein equation for the empty space, $G_{\m\n}=0$. The Liouville field $\chi(\th,\vf)$, obeying the elliptic Liouville equation in spherical coordinates (cf. \rf{Lapang})
\be
\triangle_{\th,\vf}\, \chi(\th,\vf)+2e^{\chi(\th,\vf)}-2=0\,,
\la{Phieq}
\ee
carries geometric, rather than dynamical, information on the local isothermal coordinates on a curved horizon ``surface''. Indeed, instead of the line element \rf{MinksolL} we could start from the metric
\be
d\tilde{s}^2=-f(r)dt^2+\fr{dr^2}{f(r)}+r^2 e^{\chi(x,y)}(dx^2+dy^2)\,.
\la{MinksolLf}
\ee
Then, solving for the empty space Einstein equation, we recover the Schwarzschild solution with the standard red-shift factor $f(r)=1-r_+/r$, but with the more familiar Liouville equation
\be
(\pa^2_x+\pa^2_y)\chi+2 e^\chi=0\,,
\la{LflatS2}
\ee
indicating the positive curvature of a two-dimensional manifold. Clearly, the isothermal coordinates net, determined by the Liouville field $\chi(x,y)$, can not be ``flatten'' in the plane coordinates; the trivialisation of $\chi(x,y)$ leads to the apparent contradiction. However, one can always ``flatten'' (setting $\chi=0$) the Liouville equation in spherical coordinates (cf. \rf{Phieq}), that results in turning from the quasi-spherical to the pure spherical geometry of the horizon. We conclude that the Liouville mode $\chi(\th,\vf)$ is an internal geometric characteristic of the horizon, differentiating local regions of this virtual surface from one another. This is a feature associated to the BH hair.

Now let's turn to the Eddington-Finkelstein coordinates $u$ and $v$, in terms of which the quasi-spherical Vaidya-type BH solution (cf. \rf{MinksolL}, \rf{MinksolL1}) looks as follows:
\be
ds_{in}^2=-\left(1-\fr{a(v)}{r} \right)dv^2+2dvdr+r^2e^{\chi(\th,\vf)}(d\th^2+\sin^2\th\, d\vf^2)\,,
\la{Vin}
\ee
\be
ds_{out}^2=-\left(1-\fr{a(u)}{r} \right)du^2-2dudr+r^2e^{\chi(\th,\vf)}(d\th^2+\sin^2\th\, d\vf^2)\,,
\la{Vout}
\ee
where the ingoing (advanced) null coordinate $v$ is defined by
\be
v=t+x,\qquad dx=\fr{dr}{f(r)}\equiv \fr{dr}{1-\fr{r_+}{r}}\,,
\la{EFin}
\ee
while the outgoing (retarded) null coordinate $u$ is
\be
u=t-x,\qquad dx=\fr{dr}{f(r)}\equiv \fr{dr}{1-\fr{r_+}{r}}\,.
\la{EFout}
\ee
Both metrics \rf{Vin}, \rf{Vout} are solutions to the Einstein equation with matter, $G_{\m\n}=T_{\m\n}$, where the energy-momentum tensor is that of the null dust 
\be
T^{(in)}_{\m\n}=\d_\m^v \d_\n^v\,\fr{a'(v)}{r^2},\qquad T^{(out)}_{\m\n}=-\d_\m^u \d_\n^u\,\fr{a'(u)}{r^2}\,.
\la{TmnV}
\ee
The positivity of the energy-momentum tensor (the null energy condition, NEC) requires $a'(v)\equiv da/dv >0$ and $a'(u) \equiv da/du <0$. Hence, the ingoing solution describes the process of gravitational collapse, the outgoing solution describes the emission of radiation. In any case, the integral energy density inside a surface of a constant $r$ is equal (say, for the ingoing solution) to
\be
M(v)=\int_{\th,\vf} \,d\W_2\, e^\chi \int_v \,dv \,r^2 \,\fr{a'(v)}{r^2}={a(v)} \int_{\th,\vf} \,d\W_2\, e^\chi\,.
\la{MVL}
\ee
Clearly, the total effective ``mass'' of falling into the black hole dust matter depends on the angle integral. Formally, it may be viewed as {a} result of integration of the local energy distribution over the surface of constant $r$
\be
M(v)=\int_{r=const} dS(v,\th,\vf) \,\r(v,r,\th,\vf)\,,\qquad \r=\fr{a'(v)}{r^2}\,e^{\chi(\th,\vf)}\,,
\la{MlocVL}
\ee
where exp of the Liouville field plays {a} role of the local energy distribution over the angles.

It is easy to extend the solutions \rf{Vin}, \rf{Vout} to AdS$_4$. Here we have
\be
ds_{in}^2=-\left(1-\fr{a(v)}{r}+\k^2r^2 \right)dv^2+2dvdr+r^2e^{\chi(\th,\vf)}(d\th^2+\sin^2\th\, d\vf^2)\,,
\la{VAin}
\ee
\be
ds_{out}^2=-\left(1-\fr{a(u)}{r}+\k^2r^2 \right)du^2-2dudr+r^2e^{\chi(\th,\vf)}(d\th^2+\sin^2\th\, d\vf^2)\,,
\la{VAout}
\ee
where the ingoing and outgoing null coordinates {are} defined as in \rf{EFin}, \rf{EFout} but with $dx=dr/f(r)$ for $f(r)=1-(1+\k^2 r^2_+)r_+/r+\k^2 r^2$. The line elements \rf{VAin}, \rf{VAout} solve the AdS$_4$ with the null dust matter $G_{\m\n}-3 \k^2 g_{\m\n}=T_{\m\n}$. The energy-momentum tensors of the ingoing and outgoing solutions are that in \rf{TmnV}. As in the flat case the exp of the Liouville field is the energy surface density distribution over the spherical angles.

The quasi-spherical Vaidya-type Minkowski/AdS BH solutions may be straightforwardly extended to the Bonnor-Vaidya-type black holes \cite{Bonnor:1970zz}. The electrically charged quasi-spherical BH solution of the Einstein-Maxwell equations of motion
\be
G_{\m\n}-3\k^2 g_{\m\n}-2\left( F_{\m\l}{F^{\l}}_{\n}-\fr14g_{\m\n}F^{\r\s}F_{\r\s} \right)=T_{\m\n}
\quad  \mathrm{ ( \k =0 \,\,Minkowski; \,\k \ne 0 \,\,AdS)}, 
\la{BVEin}
\ee
\be
\nabla^\m F_{\m\n}=J_\n\,,
\la{BVMax}
\ee
corresponds to
\be
ds_{in}^2=-\left(1-\fr{a(v)}{r}+\fr{q(v)^2}{r^2} \right)dv^2+2dvdr+r^2e^{\chi(\th,\vf)}(d\th^2+\sin^2\th\, d\vf^2)\,,
\la{VBin}
\ee
\be
ds_{out}^2=-\left(1-\fr{a(u)}{r}+\fr{q(u)^2}{r^2} \right)du^2-2dudr+r^2e^{\chi(\th,\vf)}(d\th^2+\sin^2\th\, d\vf^2)\,,
\la{VBout}
\ee
and
\be
A^{(in)}_{\m}=\left(\fr {q(v)} r,0,0,0\right)\,,\quad A^{(out)}_{\m}=\left(\fr {q(u)} r,0,0,0\right).
\la{VBA}
\ee
in terms of the ingoing/outgoing null coordinates $v$/$u$. The stress-energy tensor $T_{\m\n}$ and the current $J_\m$ are that of the electrically charged collapsing/radiating null dust:
\be
T^{(in)}_{\m\n}=\d^v_{\m}\d^v_{\n} \left( \fr{a'(v)}{r^2}-\fr{2q(v)q'(v)}{r^3} \right),\quad T^{(out)}_{\m\n}=-\d^v_{\m}\d^v_{\n} \left( \fr{a'(u)}{r^2}-\fr{2q(u)q'(u)}{r^3} \right),
\ee
\be
J^{(in)}_\m=\d_{\m}^v\,\fr{q'(v)}{r^2}, \quad  J^{(out)}_\m=-\d_{\m}^u\,\fr{q'(u)}{r^2}\,.
\la{Jr}
\ee
The NEC requires $r a'(v)>2q(v)q'(v)$ or $r a'(u)<2q(u)q'(u)$.

The total energy density inside a ``sphere" of the constant radius $r$  (e.g., for the ingoing solution) is equal to
\be
M(v)=\int_{\th,\vf} \,d\W_2\, e^\chi \int_v \,dv \,r^2 \,\fr{r a'(v)-(q(v)^2)'}{r^3}=\left( a(v)-\fr{q(v)^2}{r} \right) \int_{\th,\vf} \,d\W_2\, e^\chi\,,
\ee
while for the charge distribution we have
\be
Q(v)=\fr1{4\pi}\int F_{0i}dS_i=\fr1{4\pi}\int_{\th,\vf} d\W_2 \,e^{\c} \, r^2 \, \fr{q(v)}{r^2}=q(v)\,\fr1{4\pi}\int_{\th,\vf} d\W_2 e^{\c}\,.
\ee
Again, the Liouville field is responsible for the distribution both mass and charge over the angles.

\section*{Appendix B. Solutions to the Liouville equation in spherical coordinates }

\addcontentsline{toc}{section}{Appendix B. Solutions to the Liouville equation in spherical coordinates}

\def\theequation{B.\arabic{equation}}
\setcounter{equation}0

The Schwarzschild black hole solutions \rf{BHsol} with \rf{spechg} and \rf{specAdS} obey the Einstein equations in flat and AdS$_4$ space-times if the inhomogeneity distribution function $\chi(\th,\vf)$ satisfies the Liouville equation
\be
\triangle_{\th,\vf}\, \chi(\th,\vf)+2e^{\chi(\th,\vf)}-2=0\,,
\la{Leqsph}
\ee
with the angular part of the 4D Laplacian \rf{Lapang}.
In this Appendix we will establish the connection of real solutions to \rf{Leqsph} with the previously known real solutions to the Liouville equation \cite{Crowdy97}.

Our task is essentially simplified with introducing the complex coordinates on the stereographic projection plane (see \cite{PRbook87}, Section 1.2)
\be
z=e^{i\vf} \tan \left(\th/2 \right),\qquad \zb=e^{-i\vf} \tan \left(\th/2 \right)\,.
\la{zzbPP}
\ee
The two-sphere metric $ds^2=d\th^2+\sin^2\th d\vf^2$ turns into
\be
ds^2=4\,\fr{dz\,d\zb}{(1+z \zb)^2}\,,
\la{S2zzb}
\ee
that is a particular case ($n=1$) of the Fubini-Study metric for $CP^n$ complex projective space. Then the original Liouville equation \rf{Leqsph} becomes
\be
\triangle_{z,\zb} \chi(z,\zb)+2 e^{\chi(z,\zb)}-2=0\,,
\la{Leqz}
\ee
where the angular part of the 4D Laplacian is 
\be
\triangle_{z,\zb}=(1+|z|^2)^2 \pa_z \pa_{\zb}\,.
\la{Lapangzzb}
\ee
Indeed, by use of \rf{zzbPP}, $\triangle_{z,\zb}=4{\triangle_{\th,\vf}}$.

\ssk
To solve the Liouville equation \rf{Leqz} first consider an auxiliary 4D metric
\be
ds^2=-f(r) dt^2+\fr{dr^2}{f(r)}+4 r^2 e^{\chi(z,\zb)}\,dz\,d\zb \,,
\la{AdSBHint}
\ee
which is a solution to the flat/AdS$_4$ space Einstein equations with the red-shift factor \rf{specAdS} and with the distributions function $\chi(z,\zb)$ satisfying the auxiliary equation
\be
\pa_z \pa_{\zb}\,\chi(z,\zb)+2 e^{\chi(z,\zb)}=0\,.
\la{xizzbeqint}
\ee
Equation \rf{xizzbeqint} is the elliptic Liouville equation in $(z,\bar{z})$ coordinates the real solutions to which are well known (see, e.g., \cite{Crowdy97}). For example,
\be
\chi(z,\bar{z})=-2\ln \left[f(z)\bar{f}(\bar{z})+1\right]+\ln\left[f'(z) \bar{f}'(\bar{z})\right]
\la{chiexamp}
\ee
with arbitrary complex analytic function $f(z)$ solves \rf{xizzbeqint}.

Now we restore the full dependence on $z,\zb$ in the black hole background metric
\be
ds^2=-f(r) dt^2+\fr{dr^2}{f(r)}+4 r^2 e^{\xi(z,\zb)}\,\fr{dz\,d\zb}{(1+z \zb)^2} \,.
\la{AdSBHLzzb}
\ee
With $f(r)$ of \rf{specAdS} $\xi(z,\zb)$ obeys
\be
\triangle_{z,\zb}\,\xi(z,\zb)+2e^{\xi(z,\zb)}-2=0\,.
\la{Leqzzb}
\ee
It is easy to check, $\xi(z,\zb)$ relates to $\chi(z,\zb)$ via
\be
\xi(z,\zb)=\chi(z,\zb)+\ln (1+z \zb)^2 \,.
\la{xizzb}
\ee
Hence, the solution to \rf{Leqzzb} has found. With the established \rf{xizzb} it is a technical point to get from $(z,\zb)$ to the original angular coordinates back.

\section*{Appendix C. Useful relations for the Gaussian hypergeometric function}

\addcontentsline{toc}{section}{Appendix C. Useful relations for the Gaussian hypergeometric function}

\def\theequation{C.\arabic{equation}}
\setcounter{equation}0

The Gaussian hypergeometric function ${}_2 F_1[a,b,c;z]$ is defined by the following series
\be
{}_2 F_1 [a,b,c;z]=1+\fr{a\cdot b}{1\cdot c}\,z+\fr{a(a+1)b(b+1)}{1\cdot 2\cdot c (c+1)}\,z^2+\fr{a(a+1)(a+2)b(b+1)(b+2)}{1\cdot 2 \cdot 3 \cdot c (c+1)(c+2)}\,z^3+\dots
\la{2F1Series}
\ee
From this definition it is clear {that} ${}_2 F_1[a,b,c;0]=1$ for $c\ne 0$. Note also
\be
{}_2 F_1 \left[a,b,c;1\right]=\fr{\G(c) \G(c-a-b)}{\G(c-a)\G(c-b)},\quad \text{Re}(c-a-b)>0\,.
\la{Fpro1}
\ee

We use the following (see, e.g.,  \cite{BEbook},\cite{NISTbook}) relations between hypergeometric functions:
\be
\mathbf{F}[a,b,c;z]=(1-z)^{-a}\,\mathbf{F}[a,c-b,c;\fr{z}{z-1}]=(1-z)^{-b}\,\mathbf{F}[c-a,b,c;\fr{z}{z-1}],
\la{Fpro5}
\ee
\[
\mathbf{F}=\fr{{}_2 F_1 [a,b,c;z]}{\G(c)}\,,\qquad \quad |\text{arg}(1-z)|<\pi\,;
\]
and
\[
\fr{\sin(\pi (b-a))}{\pi}\, \mathbf{F} [a,b,c;z]=\fr{(-z)^{-a}}{\G(b)\G(c-a)}\, \mathbf{F}[a,a-c+1,a-b+1;\fr1{z}]
\]
\be
-\fr{(-z)^{-b}}{\G(a)\G(c-b)}\, \mathbf{F} [b,b-c+1,b-a+1;\fr1{z}]\,,\quad |\text{arg}(-z)|<\pi\,.
\la{Fpro3}
\ee

So, according to \rf{Fpro5}
\be
{}_2F_1 \left[\fr{\n+\hat{\w}}{2},\fr{3+\n+\hat{\w}}2,\fr32+\n;-\k^2 r^2\right] =(1+\k^2 r^2)^{-\fr{\n+\hat{\w}}2} {}_2F_1 \left[\fr{\n+\hat{\w}}{2},\fr{\n-\hat{\w}}2,\fr32+\n;\fr{\k^2 r^2}{1+\k^2 r^2}\right] 
\la{Fz/z-11}
\ee
and (see \rf{Fpro3})
\[
{}_2F_1 \left[-\fr{1+\n-\hat{\w}}{2},\fr{2-\n+\hat{\w}}2,\fr12-\n;-\k^2 r^2\right]=\fr{{\pi}^{1/2}}2 \fr{\G(\fr12-\n)}{\G(\fr{2-\n+\hat{\w}}2)\G(\fr{2-\n-\hat{\w}}2)}(\k^2 r^2)^{\fr{1+\n-\hat{\w}}2} \,\times
\]
\be
\times\, {}_2F_1 \left[-\fr{1+\n-\hat{\w}}{2},\fr{\n+\hat{\w}}2,-\fr12;-\fr1{\k^2 r^2}\right]
\la{Fz/z-12}
\ee
\[
+\fr4{3\pi^{3/2}}\,\cos\left(\fr{\pi}2 (\n-\hat{\w})\right)\cos\left(\fr{\pi}2(\n+\hat{\w})\right)\G\left(\fr12-\n \right)\G\left(\fr{3+\n-\hat{\w}}2\right)\G\left(\fr{3+\n+\hat{\w}}2\right)(\k^2 r^2)^{-\fr{2-\n+\hat{\w}}2} \times
\]
\[
\times\,{}_2F_1 \left[\fr{2-\n+\hat{\w}}2,\fr{3+\n+\hat{\w}}2,\fr52;-\fr1{\k^2 r^2}\right].
\]
Expressions \rf{Fz/z-11}, \rf{Fz/z-12} are used in deriving eq. \rf{Qdzone} in the main body of the paper.


\bsk\bsk\bsk

\end{document}